# Inward-growth plating of lithium driven by solid-solution based alloy phase for highly reversible lithium metal anode

Song Jin[1,2,5], Yadong Ye[1,2,5], Yijie Niu[1,3], Yansong Xu[4], Hongchang Jin[1,2], Jinxi Wang[1,2], Zhaowei Sun[1,2], Anmin Cao[4], Xiaojun Wu[1,3], Yi Luo[1], Hengxing Ji[1,2]* and Li-Jun Wan[2,4]

[1]Hefei National Laboratory for Physical Sciences at the Microscale, Hefei 230026, China.
[2]Department of Applied Chemistry, CAS Key Laboratory of Materials for Energy Conversion, University of Science and Technology of China, Hefei 230026, China.
[3]Department of Materials Science and Engineering, CAS Key Laboratory of Materials for Energy Conversion, University of Science and Technology of China, Hefei 230026, China.
[4]CAS Key Laboratory of Molecular Nanostructure and Nanotechnology, Institute of Chemistry, Chinese Academy of Sciences, Beijing 100190, China
[5]These authors contributed equally: S. Jin, Y. Ye
*email: jihengx@ustc.edu.cn

**Abstract**

Lithium metal batteries (LMB) are vital devices for high-energy-density energy storage, but Li metal anode is highly reactive with electrolyte and forms uncontrolled dendrite that can cause undesirable parasitic reactions thus poor cycling stability and raise safety concerns. Despite remarkable progress made to partly solve these issues, the Li metal still plate at the electrode/electrolyte interface where the parasitic reactions and dendrite formation invariably occur. Here we demonstrate the inward-growth plating of Li into a metal foil while avoiding surface deposition, which is driven by the reversible solid-solution based alloy phase change. Lithiation of the solid solution alloy phase facilitates the freshly generated Li atoms at the surface to sink into the foil, while the reversible alloy phase change is companied by the dealloying reaction during delithiation, which extracts Li atoms from inside of the foil. The yielded dendrite free Li anode produces an enhanced Coulombic efficiency of $99.5 \pm 0.2\%$ with a reversible capacity of 1660 mA h g$^{-1}$ (3.3 mA h cm$^{-2}$).

**Main text**

Secondary Li metal batteries (LMBs) have drawn significant attention due to their promise for enabling cell level energy density of >300 Wh kg$^{-1}$ for powering electric vehicles and portable electronics, owing to the fact that the Li metal anode has a high gravimetric capacity (3,860 mA h g$^{-1}$) and the lowest negative electrochemical potential (−3.040 V versus standard hydrogen electrode)[1,2]. However, Li dendrite formation at the Li metal surface induces serious safety concerns. Moreover, the highly negative electrochemical potential of Li/Li$^+$ can virtually reduce any electrolyte at the Li metal surface, resulting in the irreversible consumption of both Li and the electrolyte and consequently lowering the reversibility and cycling life of the battery[3,4]. In recent years, different strategies have been developed to resolve these problems, such as optimizing the electrolytes[5-9], engineering of artificial protection layers[10-13], and using nanostructured high-surface-area conductive scaffolds[14-19]. **However, alternative Li plating/stripping reactions still take place on the Li metal surface.** The renewed formation of the Li metal/electrolyte interface, when cycling a LMB, are accompanied by repeated parasitic reactions and inevitable Li dendrite formation[20]. Consequently, good cycling life of Li metal anodes were usually obtained using a flooded electrolyte and large excess of Li metal, which in turn reduces the energy density of the battery.

It was found that anode materials (e.g. Si and Sn) based on electrochemical alloying reactions allow Li-ions to transfer inside the electrode materials, which could yield a more stable electrode/electrolyte interface to be free of Li dendrites[21,22], even though the reactions take place at the potential much higher than Li/Li$^+$ redox couple and are associated with pronounced charge-discharge voltage hysteresis that sacrifice the energy of the battery. The electrochemical alloying reactions of Li can be divided into two categories: (1) reconstitution reaction and (2) solid-solution reaction[23]. The reconstitution reaction (*e.g.* Si and Sn) involves significant phase change, which requires additional activation energy that results in much higher discharge-charge voltage and hysteresis than those for Li metal plating-stripping[24]. The solid-solution reaction involves much less structure change than its counterpart in the lithiation-delithiation process, therefore can take place with a low charge-discharge voltage hysteresis at a potential that is very close to that of Li/Li$^+$ redox couple[25,26]. **When such solid-solution reaction dominates the lithiation-delithiation process, freshly generated Li atoms at the metal foil surface could be able to transfer inside the foil to formulate alloy phase rather than sitting on the surface to formulate Li metal, which should be able to avoid the Li dendrite formation and inhibit the parasitic reactions.**

In this work, we report a new type of metal anode that enables the inward-growth plating of Li into the metal foil rather than surface depositing, to inhibit parasitic reactions and avoid dendrite formation in lithiation-delithiation cycles. The alloying-dealloying reactions of a Li$_x$Ag ($x$ = 4.7 – 20) metal foil involve highly reversible phase changes of alloy at the potentials of –0.015 and +0.015 V vs. Li/Li$^+$, respectively. We found that the alloying reaction directs freshly generated Li atoms to transfer deep into the foil during the lithiation, and the reversed dealloying reaction extracts Li atoms from inside of the foil at the delithiation. Such a process completely avoids the alternative Li plating-stripping on the metal surface, thereby yields a dendrite free metal anode with inhibited parasitic reactions, which enables a LMB full-cell to work at low electrolyte consumption and anode-to-cathode capacity ratio of 1.1 for steady cycling.

**Solid-solution based reversible phase-change of Li-Ag alloy**

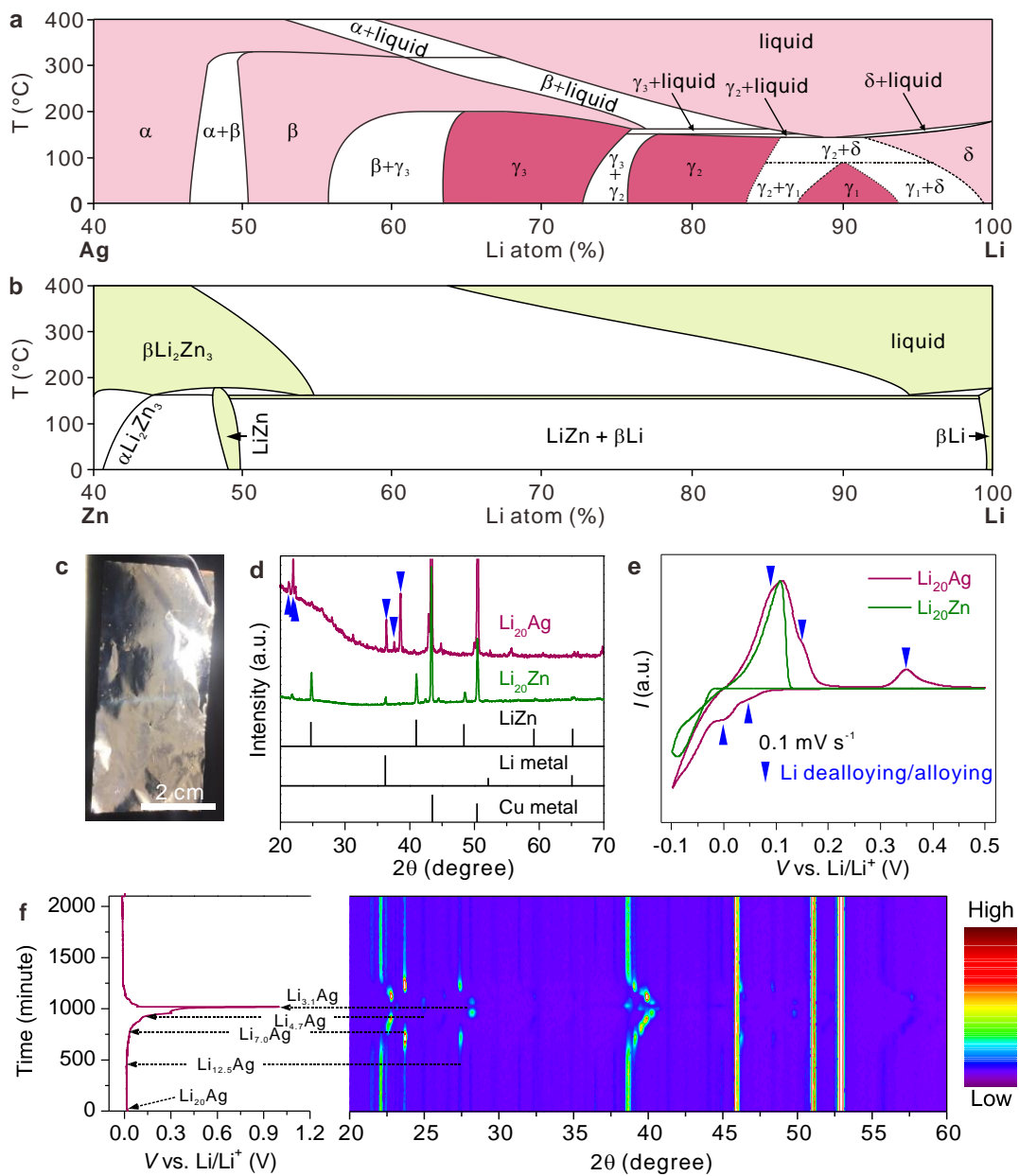

**Figure 1 | Solid-solution reaction of Li-Ag alloy.** The Li-Ag (**a**) and Li-Zn (**b**) binary phase diagrams. **c,** Photograph of Li$_{20}$Ag casted on Cu foil. **d,** XRD patterns of the Li$_{20}$Ag and Li$_{20}$Zn which are draw along with the simulated XRD patterns of LiZn intermetallic compound (Pdf No. 3-954), Li (Pdf No. 1-1131) and Cu (Pdf No. 1-1241) metals. The peaks of Cu originate from the Cu substrate. **e,** CV curves of electrochemical cells assembled with Li$_{20}$Ag or Li$_{20}$Zn as working electrode with respect to Li metal acting as both the reference and counter electrodes. **f,** In-*situ* XRD patterns of Li$_{20}$Ag anode during the second cycle of delithiation-lithiation under galvanostatic mode at a current density of 0.32 mA cm$^{-2}$. See Methods for the calculation of the atom ratios of Li/Ag in the alloy foil at different lithiation-delithiation stages.

The solid-solution reaction requires an element that forms solid-solution alloy with Li. Li-Ag alloy at Li rich composition is an attractive system (Fig. 1a), in which a solid solution with the cubic structure covers a wide range of 100 to 63 in Li atom %, namely, the δ, $\gamma_1$, $\gamma_2$, and $\gamma_3$ phases[27]. Especially, the $\gamma_1$, $\gamma_2$, and $\gamma_3$ phases have very close lattice constants of 0.980, 0.968, and 0.949 nm, respectively[28]. In contrast, the reconstitution alloying reaction occurs in intermetallic alloys. As an example in comparison to Li-Ag, Li-Zn alloy is a mixture of Li metal and intermetallic compound LiZn in a wide Li atom concentration range of 50 to 99 at.% (Fig. 1b)[29].

We obtained the $Li_{20}Ag$ and $Li_{20}Zn$ alloy foils by cooling the molten Li-Ag and Li-Zn mixture in the Li:M mole ratio of 20:1 to room temperature (see Methods). The alloy foils of ~20 cm$^2$ area can be readily fabricated (Fig. 1c). The X-ray diffraction (XRD) pattern of the $Li_{20}Zn$ alloy (Fig. 1d) presents intensive peaks at 24.7°, 41.0°, 48.4°, 59.2°, and 65.0° which match those of intermetallic compound LiZn (Pdf No. 3-954), and presents peaks at 36.1° and 65.2° which match those of metallic Li (Pdf No. 1-1131). However, the diffraction peaks of metallic Li were not observed in the XRD of $Li_{20}Ag$ alloy (Fig. 1d). Instead, the diffraction peaks at 21.3°, 22.0°, 22.3°, 36.4°, 37.6°, and 38.6° can be assigned to the γ phase Li-Ag solid-solutions. These XRD results confirm that the $Li_{20}Zn$ consists of metallic Li and intermetallic LiZn, while the $Li_{20}Ag$ is a complex solid-solution in which metallic Li is not detectable, which are in accordance with the Li-Zn and Li-Ag phase diagrams (Fig. 1a and 1b), respectively. The microstructure of the $Li_{20}Ag$ foil was presented in Supplementary Fig. S1. The cyclic voltammetry (CV) of $Li_{20}Zn$ shows typical features of an electrochemical corrosion reaction which can be assigned to the metallic Li stripping-plating (Fig. 1e)[30]. However, the cathodic and anodic peaks of dealloying and alloying reactions[31], respectively, can be observed in the $Li_{20}Ag$.

To further understand the role of solid-solution phase in the delithiation-lithiation process of the $Li_{20}Ag$, we performed in-*situ* XRD study on a two-electrode cell, which was assembled with a $Li_{20}Ag$ foil as the working electrode and a Li metal foil as both the counter and reference electrodes in an electrolyte composed of 1M lithium bis(trifluoromethanesulfonyl)imide (LiTFSI) in 1,3-dioxolane (DOL): dimethoxyethane (DME) (1:1 by volume). The cell presents an open circuit voltage of 0.0 V and the delithiation and lithiation voltages of 0.015 and –0.015 V, respectively (Fig. 1f and Supplementary Fig. S2). This delithiation-lithiation voltage profile is almost identical to that of Li stripping-plating on a Cu metal foil[32]. The XRD patterns, acquired every 30 minutes, changes with time and reveals that the Li-Ag alloy experienced four structure states in the first delithiation process (Supplementary Fig. S2). The structure of the Li-Ag alloy restores in the subsequent lithiation of the first cycle, and it changes forth and back reversibly in the next delithiation-lithiation cycles (Fig. 1f). Nevertheless, the diffraction peaks of Li metal are not observed when cycling the $Li_{20}Ag$. These *in-situ* XRD results clearly revealed that the delithiation-lithiation process of the $Li_{20}Ag$ is accompanied with the evolution of Li-Ag alloy phase. Moreover, in the delithiated state, we found that the Li/Ag atom ratio can be lowered down to 4.7 at the delithiation cutoff voltage of 0.1 V to ensure a high CE after long term cycling (Supplementary Fig. S3). It is noted here that the reversible change in the Li/Ag atom ratio of the Li-Ag alloy from 20 to 4.7 corresponds to a gravimetric capacity of 1660 mA h g$^{-1}$ (see Methods). Therefore, instead of common Li stripping-plating at the metal foil surface, the dealloying-alloying reactions should be the mechanism that the $Li_{20}Ag$ foil follows in the delithiation-lithiation cycles.

**Inward-transfer and reversible extraction of Li atoms in the Li-Ag alloy foil**

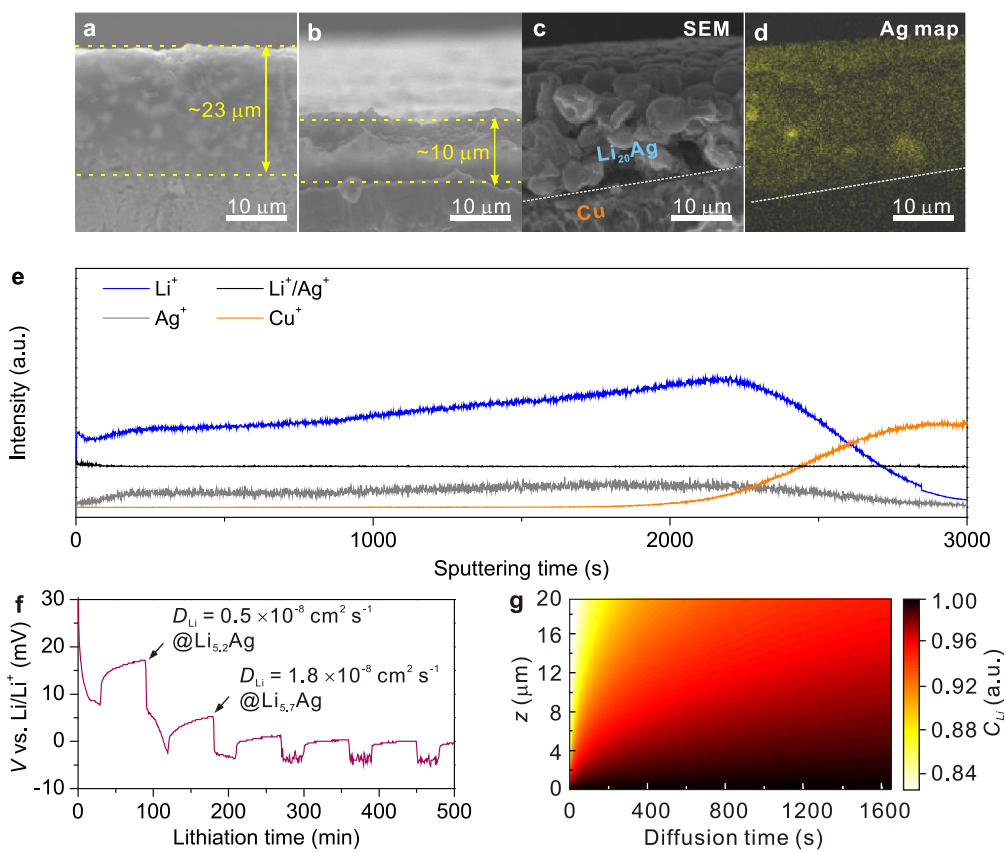

**Figure 2 | Lithiation-delithiation behavior of Li-Ag alloy.** Cross sectional SEM images of $Li_{20}Ag$ covering on Cu **(a)** and that after delithiation at 0.5 mA cm$^{-2}$ to 0.1 V vs. Li/Li$^+$ **(b)**. **c** and **d,** Cross sectional SEM image and corresponding Ag element mapping, respectively, of the $Li_{20}Ag$ anode after 20 times of delithiation-lithiation cycle. The delithiation cut-off voltage was 0.1 V vs. Li/Li$^+$. **e,** ToF-SIMS depth profiles of Li$^+$, Ag$^+$, and Cu$^+$ of a $Li_{20}Ag$ foil that undertook 20 cycles of delithiation-lithiation. **f,** GITT curve of the Li-Ag alloy applied for calculating the Li atom diffusion coefficient. **g,** Theoretically calculated the normalized Li atom concentration as a function of diffusion time and depth ($z$) in a $Li_{20}Ag$.

To study the lithiation behavior, we observed the morphology of $Li_{20}Ag$ at different cycling states. A $Li_{20}Ag$ foil of a thickness of around 23 μm (Fig. 2a) was discharged to 0.1 V vs. Li/Li$^+$ at 0.5 mA cm$^{-2}$ to allow the Li/Ag atom ratio decreasing to 4.7, which yields a foil of a thickness of around 10 μm (Fig. 2b). Then the foil was cycled for 20 times in the Li/Ag atom ratio range of 20 to 4.7, corresponding to the areal and gravimetric capacities of 3.3 mA h g$^{-1}$ and 1660 mA h g$^{-1}$, respectively. After 20 times of cycling, the foil thicknesses at the lithiated ($Li_{20}Ag$) and delithiated states ($Li_{4.7}Ag$, Supplementary Fig. S4) are almost identical to those at the first cycle, indicating a reversible thickness change of the $Li_{20}Ag$ foil. Notably, the SEM image and the corresponding EDX map (Fig. 2c and 2d) acquired from the cross-section of the Li$_{20}$Ag after 20 cycles of delithiation-lithiation show uniform Ag distribution across the foil. And the Li and Ag elements distribution across the cycled Li$_{20}$Ag foil at the lithiated state were further studied by ex-*situ* time-of-flight secondary ion mass spectrometry (ToF-SIMS). The depth profiles (Fig. 2e), which was acquired from a 50 × 50 μm area (Supplementary Fig. S5), of both the Li$^+$ and Ag$^+$ show peaks at sputtering times of 2200 s and start to decay with the simultaneous rise of Cu$^+$ (a Cu foil was applied as the substrate under the Li$_{20}$Ag, see Methods). The decay of Ag$^+$ and Li$^+$ and the rise of Cu$^+$ indicate that after Cs$^+$ milling through the Li$_{20}$Ag layer, the Cu substrate is exposed. Importantly, the intensity ratio of Li$^+$/Ag$^+$ remains almost constant (black curve in Fig. 2e), indicating the uniform distribution of Li and Ag along the vertical direction in the Li$_{20}$Ag foil after 20 times of cycling. It is reasonable to assume that if Li metal was formed and deposited at the foil surface, the Li$^+$/Ag$^+$ ratio would decay with sputtering time rather than stay constant. Moreover, in such a case, the cross sectional EDX mapping of Ag would show brighter areas at the bottom of the Li$_{20}$Ag foil. Thus, both TOF-SIMS and SEM-EDX studies confirm uniform Li and Ag distribution along the vertical direction of the cycled-Li$_{20}$Ag, which indicate the inward-transfer and reversible extraction of Li in the Li-Ag alloy foil when lithiation and delithiation, respectively.

Since the Li-Ag alloy is highly electronically conductive, the Li$^+$ from electrolyte should be reduced to formulate Li atom at the alloy surface before diffusing inside. Therefore, the diffusing of Li atom in the Li-Ag alloy is a critical process to enable the inward-transfer of Li. The diffusion coefficient of Li atom in our Li-Ag alloy measured by galvanostatic intermittent titration technique (GITT) is around 10$^{-8}$ cm$^2$ s$^{-1}$ (Fig. 2f), which is in accordance with that in Li-metal alloys reported in previous studies (10$^{-10}$ – 10$^{-6}$ cm$^2$ s$^{-1}$)[33,34] and is much higher than in bulk Li metal (5.7 × 10$^{-11}$ cm$^2$ s$^{-1}$)[35]. In this regards, we can project the normalized Li concentration, $c(z, t)$, across the alloy foil as a function of the diffusion thickness, $z$, and diffustion time, $t$, by the Fick's second law:

$$c(z, t) = c_s - (c_s - c_0)\, \text{erf}\left(\frac{z}{2\sqrt{Dt}}\right)$$

where $c_s$ is the normalized Li atom concentration at the foil surface ($t > 0$, $z = 0$) and can be fixed as 1.000 in view of the very high electron conductivity of Li-Ag alloy that Li$^+$ will accept an electron to form Li atom and deposit at the metal/electrolyte interface at the outset. $c_0$ is the normalized Li atom concentration in the interior of the alloy in the initial state ($t = 0$, $z > 0$). The fresh Li$_{20}$Ag in the initial state (lithiated state) has a Li/Ag atom ratio of 20 (Fig. 1) and thus $c_0 = 0.952$. However, in the delithiated state, we found that the Li/Ag atom ratio can be lowered down to 4.7 ($c_0 = 0.825$) to ensure a stable cycling performance (Supplementary Fig. S3). We consider the Li plating process starting on the delithiated Li-Ag alloy (Li$_{4.7}$Ag at $t = 0$, $z > 0$, and $c_0 = 0.825$). Fig. 2g plots the Li atom concentration as a function of the distance to the metal/electrolyte interface ($z$) and Li diffusion time ($t$). We see that at $z = 20$ μm, $c(z,t)$ reaches 0.900 at $t = 330$ s, and 0.952 at $t = 1650$ s. The fact that $c(z,t)$ is 0.952 at z = 20 μm indicates that the Li/Ag atom ratio at the bottom of the Li-Ag foil is restored to the initial value at this point, which suggests that the composition at the bottom of the Li-Ag foil is restored to the initial state within the time that is comparable to that of discharge. On the other hand, the inward-transfer of the Li atoms from the foil surface requires a Li concentration gradient, which exists between the Li-Ag foil surface and body (Fig. 2g). However, the Li concentration gradient should not exist in the pure Li metal foil. Therefore, the Li-Ag alloy phase enables both the higher Li atom diffusion coefficient and the Li concentration gradient to facilitate the unique inward-transfer rather than surface plating of Li at the lithiation.

**Dendrite free anode with inhibited parasitic reactions**

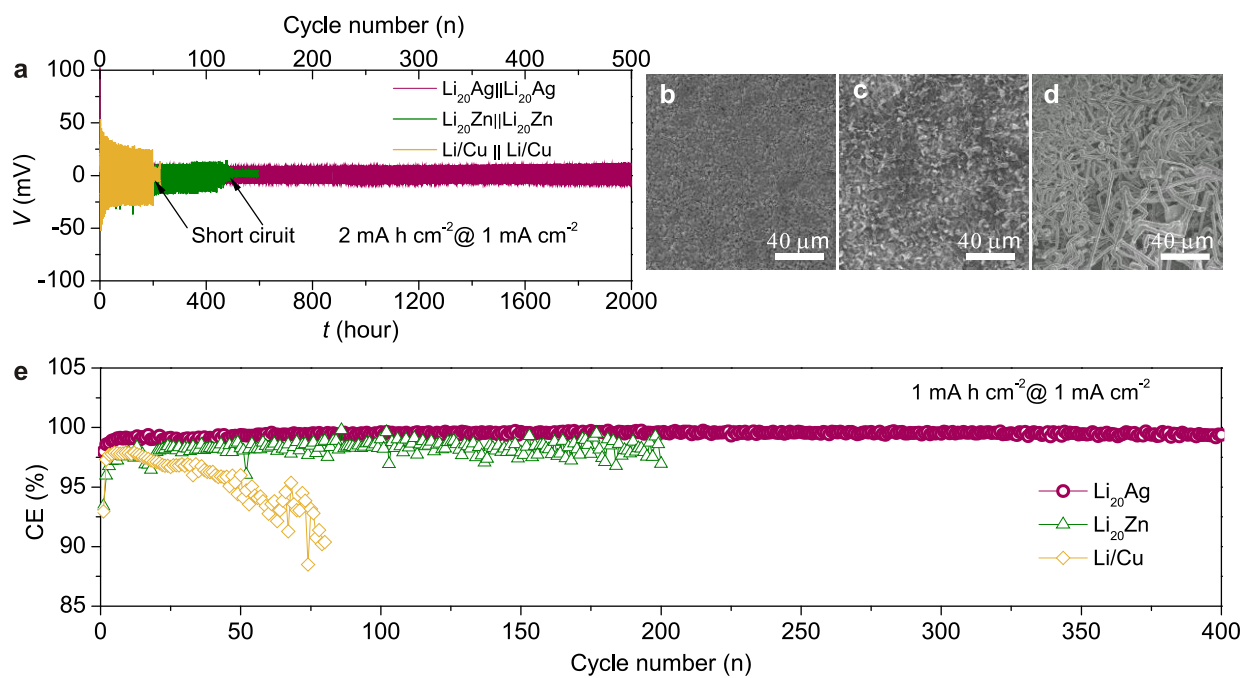

**Figure 3 | Dendrite free anode with inhibited parasitic reactions. a,** Voltage profiles of $Li_{20}Ag \parallel Li_{20}Ag$, $Li_{20}Zn \parallel Li_{20}Zn$, and Li/Cu $\parallel$ Li/Cu symmetric cells during cycling at the current density of 1 mA cm$^{-2}$. A hollow spacer having a hole at the center with diameter of 6 mm was used in the symmetric cells to accelerate the growth of Li dendrites. Morphology of $Li_{20}Ag$ **(b)**, $Li_{20}Zn$ **(c)** and Li/Cu **(d)** anodes after 20 cycles of delithiation-lithiation at 1 mA cm$^{-2}$ with areal capacity of 2 mA h cm$^{-2}$. **e,** The CE values of $Li_{20}Ag$, $Li_{20}Zn$, and Li/Cu.

The formation of Li dendrites raises safety concerns and is therefore a critical issue that hinders the wide application of Li metal anodes. To further verify the possibility of Li dendrite formation on the $Li_{20}Ag$, we assembled a symmetric cell, $Li_{20}Ag \parallel Li_{20}Ag$, with a separator having a hole at the center to allow the Li dendrite, if formed, to penetrate through. For comparison, the symmetric $Li_{20}Zn \parallel Li_{20}Zn$ cell and Li/Cu $\parallel$ Li/Cu (a Li foil was placed on Cu to form Li/Cu anode) cell were also studied. The voltage hysteresis of the $Li_{20}Zn \parallel Li_{20}Zn$ cell suddenly drops from 20 mV to 5 mV at 480 h, and that of the Li/Cu $\parallel$ Li/Cu cell suddenly drops from 50 mV to 10 mV at 200 h (Fig. 3a), indicative of short circuit induced by the eventual penetration of Li dendrites. In contrast, the $Li_{20}Ag \parallel Li_{20}Ag$ cell displays negligible voltage fluctuation which stabilize at ~20 mV at the end of more than 2000 h (500 cycles) without short circuit. This result is further confirmed by SEM images, in which the $Li_{20}Ag$ shows a dendrite-free surface (Fig. 3b), whereas Li dendrites are found on $Li_{20}Zn$ (Fig. 3c) and Li/Cu

(Fig. 3d).

The CE, a ratio of the delithiation capacity and the lithiation capacity, is a key metric to evaluate the electrochemical reversibility. These values are calculated from the voltage profiles shown in Supplementary Figs S6 – S8. As shown in Fig. 3d, the Li/Cu and $Li_{20}Zn$ anodes present average CE values of 95.4 ± 2.1% and 98.2 ± 0.7%, respectively. While the $Li_{20}Ag$ anode exhibits an average CE of 99.5 ± 0.2% for at least 400 cycles which is much higher with narrower fluctuation than the counterparts. Even in the alkyl carbonate electrolyte (1M lithium hexafluorophosphate ($LiPF_6$) in ethylene carbonate:diethyl carbonate (EC:DEC, volume ratio = 1:1)), the $Li_{20}Ag$ shows an average CE of 97.9 ± 2.0% (Supplementary Fig. S9). The high CE of $Li_{20}Ag$ indicates limited parasitic reactions such as electrolyte decomposition and Li consumption.

**Electrochemical performance of $Li_{20}Ag$ anode**

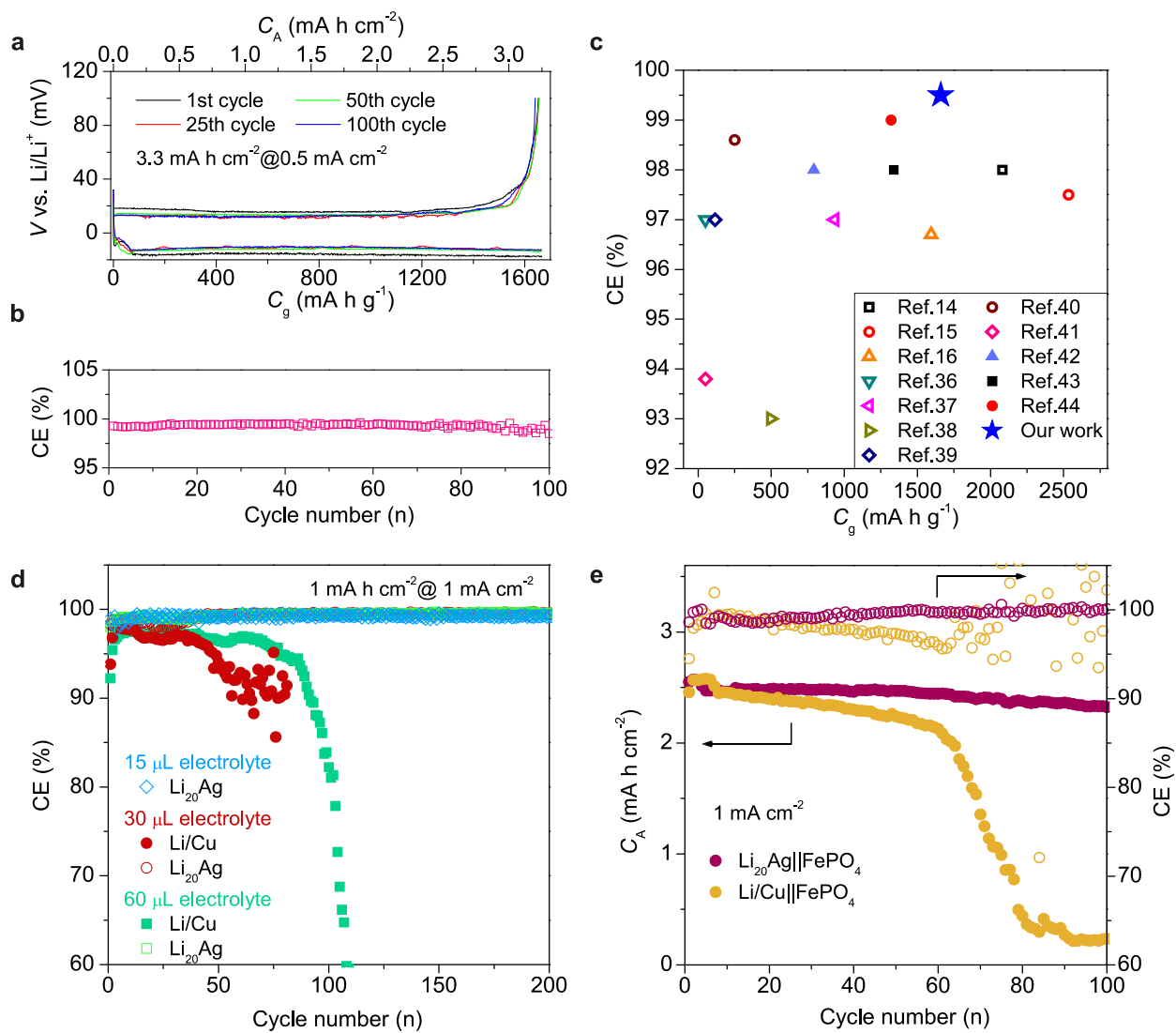

**Figure 4 | Electrochemical performance of Li$_{20}$Ag anode. a,** Voltage profiles of the 1$^{st}$, 25$^{th}$, 50$^{th}$, and 100$^{th}$ cycle of the Li$_{20}$Ag anode, from which, the CE values in panel **b** were calculated. **b,** CE of Li$_{20}$Ag anode measured at 0.5 mA cm$^{-2}$. **c,** Comparison of the gravimetric capacity and CE metrics of the Li$_{20}$Ag anode with those of various Li metal anodes. **d,** CE of Li$_{20}$Ag and Li/Cu measured with different volumes of electrolytes. **e,** Cycling performance of Li$_{20}$Ag and Li/Cu when paired with a delithiated-LiFePO$_4$ cathode (FePO$_4$) at 1 mA cm$^{-2}$. The anode/cathode capacity ratios were 1.1 in the full-cells.

Such Li$_{20}$Ag anode can offer stable electrochemical cycling with the combination of considerably high reversibility and capacity. Fig. 4a shows that the Li$_{20}$Ag delivers a reversible gravimetric and areal capacities of 1660 mA h g$^{-1}$ (normalized to the mass of Li$_{20}$Ag) and 3.3 mA h cm$^{-2}$, respectively, with an average CE of 99.5 ± 0.2% (Fig. 4b). These CE values are calculated from the voltage profiles shown in Fig. 4a, which were obtained on a Li$_{20}$Ag foil with delithiation cutoff voltage of 0.1 V. The combination of high reversible capacity and CE is critical to practical application, and our $Li_{20}Ag$ shows superior combination of gravimetric capacity and CE to state-of-the-art Li metal anodes (Fig. 4c and Supplementary Table S1)[14-16,36-47].

Results in the previous sections have shown that a high CE indicates limited parasitic reactions such as electrolyte decomposition. To further verify this effect, we have recorded the electrochemical performance of $Li_{20}Ag$ for three different electrolyte volumes of 15, 30, and 60 μL. The $Li_{20}Ag$ anodes were cycled for more than 200 times, and the CE of the cells with 15, 30, and 60 μL of electrolyte are 99.2 ± 0.1%, 99.5 ± 0.1%, and 99.6 ± 0.1%, respectively (Fig. 4d and Supplementary Fig. S10). In contrast, the CE of the Li/Cu anodes drop rapidly. In addition, the high CE also allows for LMBs to work at a low anode-to-cathode capacity ratio, which is critical to achieving high energy density at the cell level. For example, commercial LIBs contain graphite anodes with anode-to-cathode capacity ratio of (1.05 – 1.1)[48,49], which is significantly lower than the value expected for Li metal anode (1.5 – 3)[18,50]. We assembled $Li_{20}Ag$||$FePO_4$ cell with an anode-to-cathode capacity ratio of 1.1. The areal capacity of the delithiated-$LiFePO_4$ ($FePO_4$) cathode was measured to be 2.5 mA h cm$^{-2}$ by delithiation of a commercial $LiFePO_4$ electrode (Figure S11a). Then a $Li_{20}Ag$ foil of areal capacity of 2.8 mA h cm$^{-2}$ was prepared, and its areal capacity was verified by delithiating the $Li_{20}Ag$ foil (Figure S11b) to ensure the anode-to-cathode capacity ratio to be 1.1. The $Li_{20}Ag$||$FePO_4$ cell delivered an areal capacity of 2.5 mA h cm$^{-2}$ with a very high initial CE of 98.6% at the 1$^{st}$ cycle, and retained an areal capacity of 2.3 mA h cm$^{-2}$ after 100 cycles (Figure 4e and Supplementary Fig. S12). However, the Li/Cu||$FePO_4$ cell with anode-to-cathode capacity ratio of 1.1 shows scattered CE and rapid capacity decay when cycling the battery.

The $Li_{20}Ag$ can also be introduced into 3D-carbon materials by a melt-infusion process similar to Li metal (Supplementary Fig. S13 – S17). The $Li_{20}Ag$@CF paper shows a reversible areal capacity of 12 mA h cm$^{-2}$ at the current density of 1 mA cm$^{-2}$ for at least 1200 h without capacity decay. It also shows steady voltage profiles when cycled at high current densities of 2, 4, and 8 mA cm$^{-2}$, and delivers areal capacities of 2, 4, and 8 mA h cm$^{-2}$, thus presenting good rate capability. The Li$_{20}$Ag@CF paper indicating the viability of Li$_{20}$Ag for application in LMBs, also in cases where 3D current collectors are required.

**Conclusion**

In summary, the electrochemical dealloying-alloying process based on reversible solid-solution reaction can take place at the potential very close to that of Li/Li$^+$ redox couple, which is benefited from flexible structure change of the Li-metal alloy phases during lithiation-delithiation processes. The changeable Li-metal alloy phases, such as Li$_x$Ag ($x$ = 4.7 – 20) instead of metallic Li, are reversibly generated. The solid-solution based alloy phase provides not only a Li concentration gradient between the alloy phase in the foil body and the Li ad-atoms at the foil surface, but also a higher Li atom diffusing coefficient than the bulk Li metal. In this manner, the Li atoms generated at the alloy/electrolyte interface during the lithiation process can diffuse inside of the foil to formulate alloy, and during the delithiation process the Li atoms generated by dealloying can be extracted from the foil within the time that is comparable to that of discharge. This unique inward-growth plating of Li avoids dendrite formation and enables stable cycling performance with high CE, due to which, reduced electrolyte usage and low anode-to-cathode capacity ratio can be achieved in a full-cell. Our study thus describes a new mode of Li storage in metal anodes and proposes novel strategies to optimize metal anodes for next generation high-energy LMBs.

**Methods**

**Preparation of the Li$_{20}$Ag anode.** In a typical procedure, a Li foil (140 mg, Alfa Aesar) was first placed in a stainless steel crucible. The crucible was then heated on a hot plate to ~300 °C. After Li was melted, 108 mg of sliver powder (Alfa Aesar) was added into the crucible and the mixture was vigorously stirred to achieve a homogeneous reaction. Once the reaction was completed, the as-obtained alloy was cast on a copper foil with a spreader followed by cooling to room temperature in argon atmosphere. The same process was used to obtain Li$_{20}$Zn. The atomic ratio of Li foil to the metal powders in Li$_{20}$M (M=Ag and Zn) was 20:1.

**Preparation of Li$_{20}$Ag@CF paper.** Carbon fiber (200 mg, Carbon Paper 060, bulk density 0.44 g cm$^{-3}$, porosity 80%, and thickness 0.2 mm) was refluxed with a mixture of nitric acid (25 mL, 70% HNO$_3$) and sulfuric acid (75 mL, 98% H$_2$SO$_4$) at 60 °C for 5 hours. The sample was then rinsed in DI water and ethanol before drying at 50 °C in a vacuum oven. Next, 100 mg of the acid-treated carbon fiber was immersed in 100 mL of 0.1 M potassium permanganate (KMnO$_4$) solution at 50 °C for 15 minutes, whereupon, MnO$_2$ was deposited directly onto the carbon fiber substrate due to the in-situ redox reaction between KMnO$_4$ and carbon fiber. The carbon fiber was then rinsed in DI water and ethanol before drying at 50 °C. Li$_{20}$Ag melt-infusion was carried out in an argon-filled glovebox with oxygen level < 0.1 ppm. In a typical process, the edge of the carbon fiber was contacted with molten Li$_{20}$Ag. The Li$_{20}$Ag was pulled up to wet the whole matrix to form the composite electrode.

**Structural characterization.** SEM images were acquired using SIRION200 (FEI Ltd.) operated at 5.0 kV, and HRTEM images were obtained using a JEM-2100F (JEOL Ltd.) at an accelerating voltage of 200 kV. XRD were recorded (D/max-TTR III) with Cu Kα radiation of λ = 1.54178 Å operating at 40 kV and 200 mA. Elemental analysis and depth profiles of the Li$_{20}$Ag anode were conducted using time-of-flight secondary ion mass spectrometry (TOF-SIMS, TOF.SIMS5 IONTOF GmbH). The experiments were performed in static mode where the sputtering gun ($Cs^+$) was operated over a 50 × 50 μm$^2$ area of the electrode surface. Secondary ions were detected in positive ion mode.

**In-Situ XRD measurement.** In situ X-ray diffraction was carried out using a Bruker D8 Advance diffractometer with Cu Kα radiation (λ = 1.54178 Å). A specially designed Swagelok cell was equipped with a beryllium window for X-ray penetration. The Li$_{20}$Ag was cast onto the center of the beryllium window with a diameter of ~2.0 cm. Each scan took approximately 30 minutes. The in-*situ* cells were cycled at a current density of 0.32 mA cm$^{-2}$ during the charge/discharge cycles.

**Electrochemical measurements.** Electrochemical measurements were performed using CR2032-type coin cells of a two-electrode configuration. Cells were assembled by using metallic Li foil (Alfa Aesar, thickness of 700 μm, diameter: 12 mm) as the counter electrode and Li$_{20}$Ag as the working electrode (1 cm × 0.5 cm or 1 cm × 1 cm). To standardize the testing, 60 μL of electrolyte was used in each coin cell, except the CE test of Li$_{20}$Ag anode at different electrolyte volumes in the main text. Cells used to investigate dendrite growth were assembled with hollow spacers made of glass fiber (Whatman, GF/D) with a hole (diameter: 6 mm) at the center to accelerate the growth of lithium dendrites. Other cells used for measuring CE, gravimetric and areal capacities, and cycling life were assembled with PP separators (Celgard 2400). Two types of electrolytes were used: (i) 1 M LiPF$_6$ in 1:1 (v/v) EC and DEC and (ii) 1 M LiTFSI in 1:1 (v/v) DOL/DME electrolytes. Battery cycling data were collected using a LAND electrochemical testing system at 25 °C. The electrolyte used in the galvanostatic charge/discharge and CV tests was 1 M LiTFSI in 1:1 (v/v) DOL/DME electrolyte. CVs were collected at the scan rate of 0.1 mV s$^{-1}$ in the potential ranges of −0.1 to 0.5 V with Li metal as both the reference and counter electrodes. Since freshly prepared Li$_{20}$Ag or Li$_{20}$Zn foil is in the lithiated state with pre-stored Li, to measure the CE of the Li$_{20}$Ag, a pre-delithiation process was first performed. This pre-delithiation process was carried out at the constant current of 0.5 mA cm$^{-2}$ to reach a voltage of 0.1 V, which yielded an areal capacity. This value was set as the Li plating capacity in each of the subsequent lithiation-delithiation cycles, and was achieved by plating Li at a constant current. After the lithiation process of each cycle, delithiation was performed until the cell voltage reached 0.1 V to obtain the delithiation capacity. The CE of each cycle is then calculated by dividing the delithiation capacity by the lithiation.

The Li$^+$ diffusion coefficient was calculated using GITT based on the following equation:

$$D_{Li}^+ = \frac{4}{\pi\tau}\left(\frac{n_m V_M}{S}\right)^2 \left(\frac{\Delta E_s}{\Delta E_t}\right)^2$$

where $n_m$ and $V_M$ are the molar mass (mol) and volume (cm$^3$ mol$^{-1}$) of the Li$_{20}$Ag, respectively, $S$ is the electrode surface area, $\tau$ is the time duration of the current pulse, $\Delta E_t$ is the voltage drop during the discharge, and $\Delta E_s$ is the voltage change during the current pulse. For the Li$_{20}$Ag electrode, the $n_m V_M = V$, so the equation becomes

$$D_{Li}^+ = \frac{4}{\pi\tau} L^2 \left(\frac{\Delta E_s}{\Delta E_\tau}\right)^2$$

where $L$ is the thickness of the Li$_{20}$Ag foil.

To study the electrochemical performance of Li$_{20}$Ag anode in a full cell, delithiated-LiFePO$_4$ was used as the cathode. The LiFePO$_4$ cathode was purchased from MTI, which consists of LiFePO$_4$ (mass loading of 16 mg cm$^{-2}$), super P carbon, and polyvinylidene fluoride at a weight ratio of 90:5:5. The fresh LiFePO$_4$ electrode (areal capacity of ~2.5 mA h cm$^{-2}$, Figure S11a) was paired with Li foil and was charged (delithiated) to 4.2 V. After Li was completely extracted from the LiFePO$_4$, the delithiated-LiFePO$_4$ was disassembled in the glovebox and served as the cathode to pair with Li$_{20}$Ag anode (areal capacity of 2.8 mA h cm$^{-2}$, Figure S11b) for full cell testing.

**Calculation of the theoretical gravimetric capacity of Li$_x$Ag when charged to Li$_y$Ag**

When charging the Li$_x$Ag to Li$_y$Ag, which means that the number of the lithium atom extracted from the fresh Li$_x$Ag is (*x-y*), since the theoretical specific capacity of Li metal is 3860 mA h g$^{-1}$, the theoretical capacity of Li$_x$Ag when charged to Li$_y$Ag can be calculated to be:

$$\text{Theoretical capacity} = 3860 \text{ mA h g}^{-1} \times \frac{(x-y) \times M_{Li}}{x \times M_{Li} + 1 \times M_{Ag}} \times 100\%$$

$M_{Li}$ and $M_{Ag}$ are the atomic weights of lithium and sliver.

For example, the theoretical gravimetric capacity of Li$_{20}$Ag alloy when charged to Li$_{4.7}$Ag is calculated to be

$$3860 \text{ mA h g}^{-1} \times \frac{(20-4.7) \times M_{Li}}{20 \times M_{Li} + 1 \times M_{Ag}} \times 100\% = 1661 \text{ mA h g}^{-1}$$

## Acknowledgements

We appreciate funding support from the Natural Science Foundation of China (51761145046 and 21975243), support from the 100 Talents Program of the Chinese Academy of Sciences.

## Author contributions

H.X.J. and L.J.W. coordinated the project and designed the experiment. S.J. and Y.D.Y. prepared all the samples, S.J., Y.D.Y., H.C.J., Z.W.S., and J.X.W. carried out the electrochemical measurements. Y.S.X. and Y.D.Y carried out the in-*situ* XRD measurements. S.J., Y.D.Y., and Y.J.N. carried out data analysis. S.J., H.X.J., and Y.L. wrote the manuscript. All authors discussed the results and commented on the manuscript.

## Competing financial interests

The authors declare no competing financial interests.

## Additional information

Supplementary information is available for this paper online.
Reprints and permissions information is available at www.nature.com/reprints.
Correspondence and requests for materials should be addressed to H.X.J.

## Data availability

The data that support the plots within this paper and other findings of this study are available from the corresponding author upon reasonable request.